\documentclass[aps,pre,showpacs,twocolumn,amsmath,groupedaddress]{revtex4-1}

\usepackage{graphicx,amssymb,bbm,bm}

\def\ii{ {\rm i} }
\def\dd{ {\rm d} }

\def\Z{{\mathbb{Z}}}
\def\R{{\mathbb{R}}}
\def\T{{\mathbb{T}}}

\def\>{\rangle}
\def\<{\langle}

\def\tr{{{\rm tr}}}

\newcommand{\ve}[1]{{\bm #1}}
\newcommand{\op}[1]{\hat{#1}}

\begin{document}

\title{Complexity and non-separability of classical Liouvillian dynamics}

\author{Toma\v z Prosen$^{1,2}$}
\affiliation{$^1$ Department of Physics, Faculty of Mathematics and Physics,
  University of Ljubljana, Ljubljana, Slovenia \\
$^2$ Department of Physics and Astronomy, University of Potsdam, Potsdam, Germany}
 
\date{\today}

\begin{abstract}
We propose a simple complexity indicator of classical Liouvillian dynamics, namely the separability entropy,
which determines the logarithm of an effective number of terms in a Schmidt decomposition of phase space
density with respect to an arbitrary fixed product basis.
We show that linear growth of separability entropy provides stricter criterion of complexity
than Kolmogorov-Sinai entropy, namely it requires that dynamics is exponentially unstable, non-linear and non-markovian.
\end{abstract}


\maketitle

\section{Introduction}

How can one characterize algorithmic complexity of Liouville evolution 
$\dd\rho^t/\dd t = \{\rho^t,H \}_{\rm Poisson\, bracket}$
of conservative classical dynamics with Hamiltonian $H$? 
Is Kolmogorov-Chaitin complexity of individual orbits related to complexity of field solutions $\rho^t(\ve{z})$ ($\ve{z}$ denoting a collection of $2d$ phase space coordinates) of Liouville equation? The answer is `no', as shown by a paradigmatic example of chaotic dynamics, the stretching and folding baker's map, which is equivalent to the Bernoulli shift on an infinite binary symbol sequence (coin tossing), so its orbit dynamics is algorithmically complex, but its Liouville evolution is exactly solvable 
\cite{gaspard}. More generally, one can identify two extreme cases of exact solvability in conservative (closed and noise-less) classical dynamics: namely (i) orbit-wise exact solvability which is associated with 
a Liouville {\em integrability} and existence of a complete set of constants of motion, and (ii) field-wise exact solvability which is associated with an existence of a finite Markov partition and symbolic dynamics 
\cite{predrag,symbolic}.
 
The fundamental question that we address in this paper is whether the notion of complexity qualitatively changes when we focus our attention from individual orbits to time-dependent statistical ensembles? The latter is more common and meaningful in statistical mechanics. We propose to apply a concept of a Schmidt rank and entanglement entropy - common in quantum information theory \cite{eisert} -  
to a joint probability distribution of several classical (dynamical) variables, in order to describe the growth rate of complexity of description
of classical field solutions of the Liouville equation. In this way, a new and conceptually very simple measure of complexity is defined, {\em the separability complexity}, which exactly vanishes in both cases (i,ii) of exact solvability and thus hopefully detects genuinely hard cases of classical deterministic dynamics, even in the statistical sense.
 The utility of the new measure is demonstrated and compared to the characteristics of the transport and diffusion in the Fourier space (being common measures of Hamiltonian turbulence) for several non-trivial examples of chaotic and regular 2D and 4D classical dynamical maps. One should note that introducing either {\em classical noise} or {\em quantum effects} introduces a natural cutoff scale to a phase space resolution and thus qualitatively reduces such a notion of  complexity. Quantum, or noisy classical dynamics can become genuinely complex only in the (thermodynamic) limit of increasingly many degrees of freedom.

\section{Separability entropy and complexity indicators}

In order to make our discussion simple but general we shall consider discrete dynamical systems, say stroboscopic or Poincar\' e maps of
Hamiltonian dynamics, given in terms of a Lebesgue-measure preserving invertible map $\ve{z}_{t+1}=\ve{\phi}(\ve{z}_t)$ over a compact phase space 
${\cal M} \subset \R^{2d}$. The map induces a {\em unitary} Perron-Frobenius operator over the Hilbert space $L^2({\cal M})$ of phase space densities, $(\op{U} \rho)(\ve{z}) \equiv \rho(\ve{\phi}^{-1}(\ve{z}))$. For simplicity we shall identify the phase space with $2d-$dimensional torus ${\cal M}=\T^{2d}$ (while more general cases can be treated with obvious modifications) and
consider an arbitrary phase space decomposition ${\cal M} = \T^d \oplus \T^d \ni \ve{z}\equiv (\ve{x},\ve{y})$ into two sets of $d$ coordinates, which could for example, describe two disjoint subsets of degrees of freedom, or $\ve{x}$ could be positions and $\ve{y}$ momenta, etc. The phase space decomposition induces factorization of the Hilbert space of densities
$L^2({\cal M}) = L^2(\T^d)\otimes L^2(\T^d).$ Let us write time-evolved Liouville density as 
$\rho^t(\ve{x},\ve{y}) = (\op{U}^t \rho^0)(\ve{x},\ve{y})$, and normalize it in $L^2$ sense as $\int \dd^{2d}\ve{z} |\rho^t(\ve{z})|^2 = 1$. 
Then we write the Schmidt (or {\em singular value}) {\em decomposition} of the density
\begin{equation}
\rho^t(\ve{x},\ve{y}) = \sum_n v^t_n(\ve{x}) \mu^t_n w^t_n(\ve{y}),
\label{eq:decomp}
\end{equation}
in terms of two sets of {\em ortho-normalized} functions $\{v^t_n\}$, $\{w^t_n\}$, $n=1,2\ldots$ and
a set of Schmidt coefficients $\{\mu^t_1 \ge \mu^t_2 \ge \ldots \ge 0\}$ satisfying $\sum_n |\mu^t_n|^2 = 1$. In practice, we can treat $\rho^t(\ve{x},\ve{y})$ as a matrix
of row $\ve{x}$ and column $\ve{y}$ and consider sufficiently fine discretization of continuous variables $\ve{x},\ve{y} \in \T^d$ that the
results don't depend on it.
Let us define a {\em separability entropy} (s-entropy) as a logarithm of an effective number of terms in decomposition (\ref{eq:decomp})
\begin{equation}
h[\rho^t] = -\sum_n |\mu^t_n|^2 \ln |\mu^t_n|^2
\label{eq:vonneumann}
\end{equation}
which gives a quantitative measure of separability of phase-space density with respect to a given phase 
space decomposition. Alternatively, $h[\rho^t]$ can be computed as von Neuman entropy $h[\rho^t] = -\tr[R^t\ln R^t]$, where
$R^t$ are trace-class, positive, self-adjoint operators on $L^2(\T^d)$ with integral kernels
$R^t(\ve{x},\ve{x}') = \int \dd^d\ve{y} \rho^t(\ve{x},\ve{y}) \rho^t(\ve{x}',\ve{y}).$
Note that s-entropy $h[\rho^t]$ does {\em not} depend on the coordinate system we use for each phase-space
factor space, since decomposition (\ref{eq:decomp}) is invariant under invertible measure preserving 
transformations of the form 
$(\ve{x},\ve{y}) \to (\ve{\chi}(\ve{x}),\ve{\eta}(\ve{y}))$, namely it only depends on phase-space de-composition and
dynamics $\ve{\phi}$.
Any nontrivial decomposition can be generated from a canonical one in terms of some phase-space
diffeomorphism $\ve{\pi}:{\cal M}\to{\cal M}$, namely $\ve{z} = \ve{\pi}(\ve{x},\ve{y})$, and the corresponding s-entropy
is computed as $h[\rho^t \circ \ve{\pi}]$.
Now, let us assume that for sufficiently complex dynamics s-entropy can grow proportionally with time, and define its asymptotic growth rate as {\em s-complexity}
\begin{equation}
C_{\rm s}[\ve{\phi}] = \inf_{\pi} \sup_{\rho^0} \lim_{t\to\infty} \frac{1}{t} h[\rho^0 \circ \ve{\phi}^{-t} \circ \ve{\pi}]
\label{eq:scomplex}
\end{equation}
with, taking first a supremum over initial densities $\rho^0$, and later an infimum over the phase-space decompositions $\ve{\pi}$ \cite{note}.
Clearly, {\em any} complete and accurate (numerical) representation of phase space density $\rho^t$ 
needs at least ${\cal O}(\exp(h[\rho^t]))$ terms of the form (\ref{eq:decomp}), so ${\cal O}(\exp(C_{\rm s} t))$
estimates \cite{foot1} the necessary amount of classical computing resources needed to simulate Liouville dynamics
up to time $t$, but is it {\em sufficient}?

\begin{figure}[h!]
\includegraphics[width=1.025\linewidth]{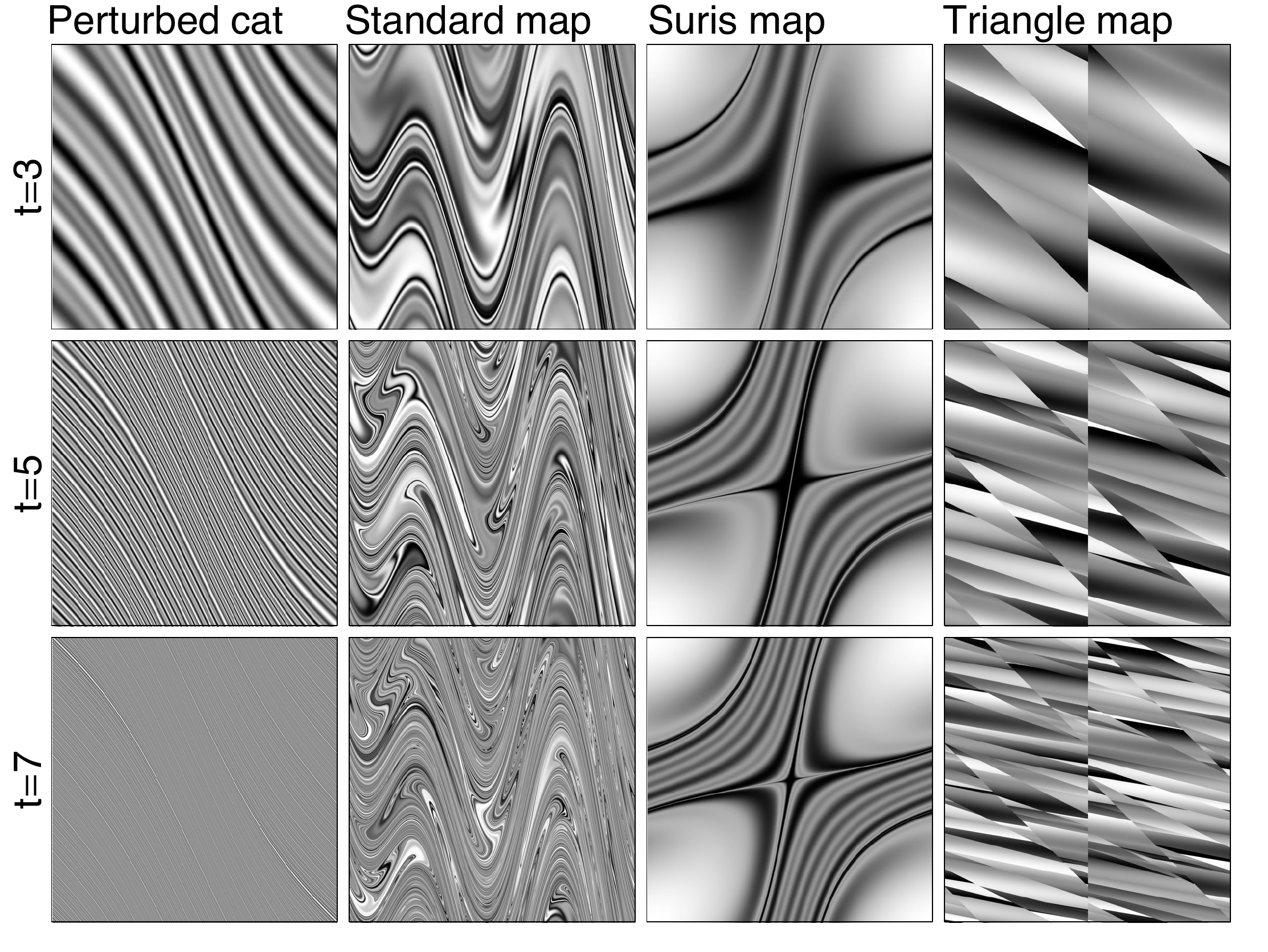}
\caption{Snapshots at $t=3,5,7$ (top-down) of Liouville dynamics starting from initial density $\rho^{t=0}(x,y) = (2 + \cos x + \cos y)(2\pi\sqrt{5})$, for the four 2D toral maps (PC, SM, IM, TM, left-right) introduced in the text. The grayscale indicates the probability density $\rho^t(x,y)$ (zero=white, maximal=black).}
\label{fig:snapsr} 
\end{figure}

\begin{figure}[h!]
\includegraphics[width=1.025\linewidth]{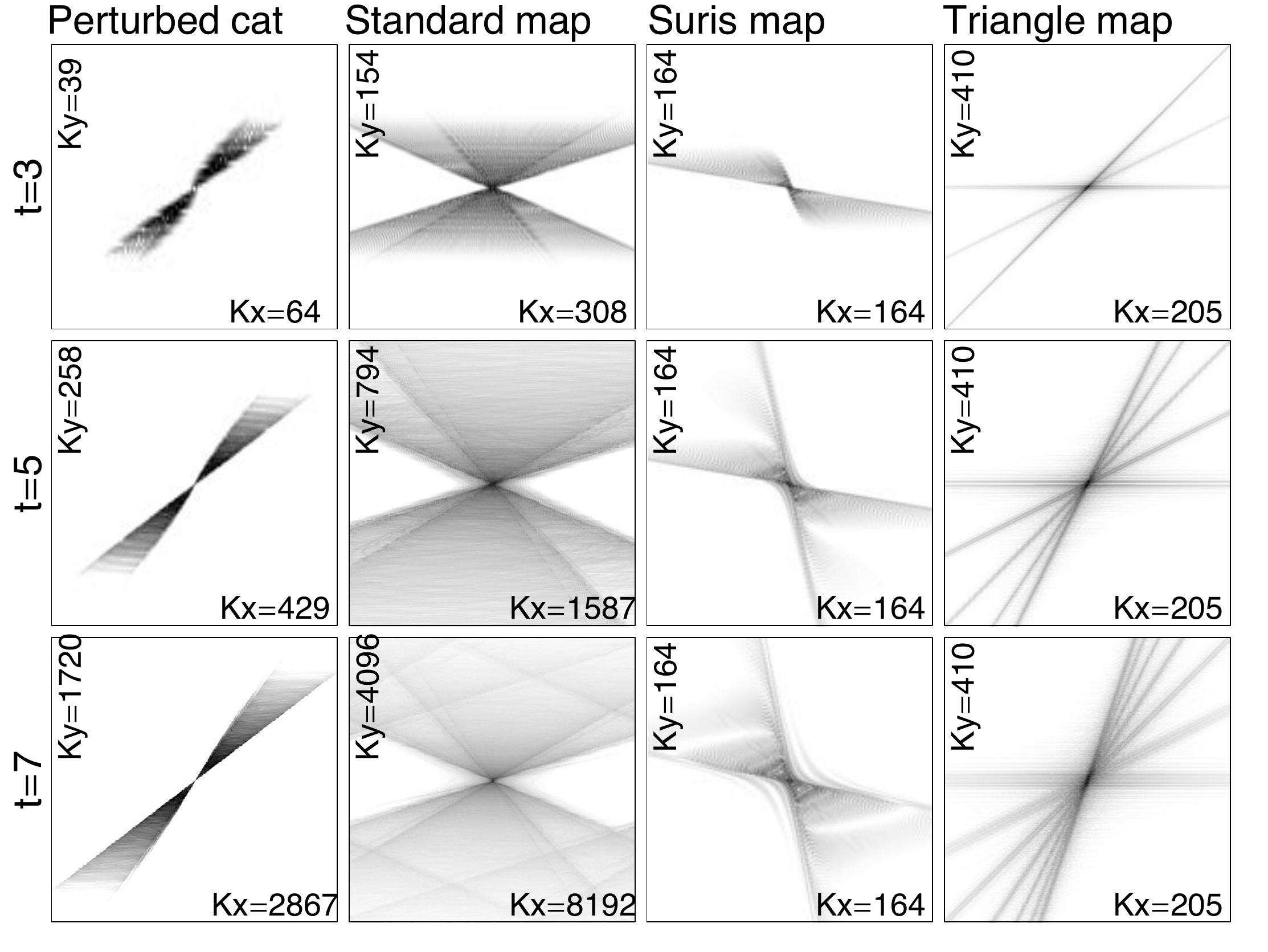}
\caption{Snapshots at $t=3,5,7$ (top-down) of Liouville density in Fourier space $\tilde{\rho}^t(k_x,k_y)$, starting from $\tilde{\rho}^0(k_x,k_y) \propto 4 \delta_{k_x,0} \delta_{k_y,0} + \delta_{|k_x|,1}\delta_{k_y,0}+
\delta_{k_x,0}\delta_{|k_y|,1}$, for the four different 2D toral dynamics (PC, SM, IM, TM, left-right) introduced in the text.
 The grayscale indicates probability density (zero=white, maximal=black), while axes labels $K_x$ and $K_y$ indicate the Fourier space range $[-K_x,K_x]\times [-K_y,K_y]$ which is scaled (both, in $x$ and $y$ direction) with the map's Lyapunov exponent $\exp(t \lambda_{\rm max})$ from top to bottom panels.}
\label{fig:snapsf} 
\end{figure}

As an alternative measure of algorithmic complexity of Liouville dynamics we define the {\em Fourier entropy} (f-entropy), as the
logarithm of an effective number of Fourier harmonics $\tilde{\rho}^t(\ve{k}) \equiv (2\pi)^{-2d}\int\dd^{2d}\ve{z} e^{\ii \ve{k}\cdot\ve{z}}\rho^t(\ve{z})$,
$\ve{k}\in\Z^{2d}$, needed to simulate the solution for time $t$, namely
\begin{equation}
g[\rho^t] = -\sum_{\ve{k}\in\Z^{2d}} | \tilde{\rho}^t(\ve{k})|^2 \ln|\tilde{\rho}^t(\ve{k})|^2,
\end{equation}
and the corresponding {\em f-complexity} as 
\begin{equation}
C_{\rm f}[\ve{\phi}]=\sup_{\rho^0} \lim_{t\to\infty}g[\rho^0\circ \ve{\phi}^{-t}]/t.
\label{eq:fcomplex}
\end{equation}
${\cal O}(\exp(C_{\rm f} t))$ gives a {\em sufficient} amount of classical computing resources needed
for accurate simulation of Liouville dynamics up to time $t$, but is it {\em necessary}?
Summarizing, we state the following two inequalities:
\begin{equation}
C_{\rm s} \le C_{\rm f} \le 2d \lambda_{\rm max}
\label{eq:ineq}
\end{equation}
where $\lambda_{\rm max}$ is the maximal Lyapunov exponent which determines the smallest
scale $\sim\exp(-\lambda_{\rm max}t)$ on which $\rho^t(\ve{z})$ can vary, in each of $2d$ phase space directions.

\begin{figure}[h!]
\includegraphics[width=0.95\linewidth]{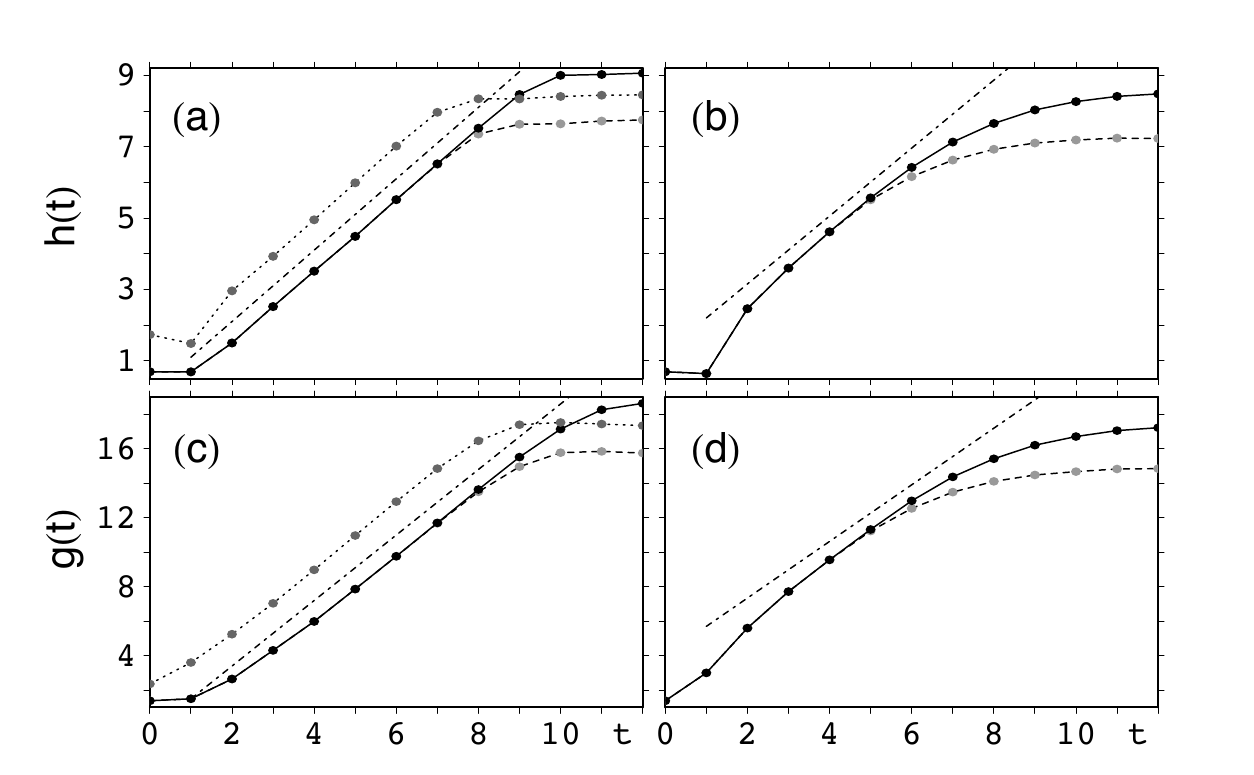}
\caption{Separability entropy $h(t)=h[\rho^t]$ (a,b) and Fourier entropy $g(t) = g[\rho_t]$ (c,d) for two cases of chaotic dynamics, PC (a,c) and SM (b,d). Discretization/trucation with $N=2^p$ nodes in real/Fourier-space along each ($x$ and $y$) direction is used, and data
for $p=14$ (black, full curves/symbols), $p=13$ (dark grey, short dash), and $p=12$ (light grey, long dash) are shown. For $p=12,14$ we use the same initial
density $\rho^0$ as in Figs.\ref{fig:snapsr},\ref{fig:snapsf}, while for $p=13$ (only for PC) a different initial state with Fourier harmonics populated up to $|\ve{k}|=4$ is used in order to demonstrate the same asymptotic growth rates, indicated with dash-dotted lines: $C_{\rm s} = 1.00$ (a), $C_{\rm s} = 0.952$ (b), $C_{\rm f} = 2\lambda_{\rm max}^{\rm PC}=1.90$ (c), $C_{\rm f} = 2\lambda_{\rm max}^{\rm SM} =  1.64$.}
\label{fig:eech} 
\end{figure}

For the first inequality (\ref{eq:ineq}) to be saturated it would mean that both, s-complexity and f-complexity yield sufficient and necessary amount of computing resources for Liouvillian simulation. As indicated later in numerical
experiments, this may not generally be true.
Fo the second inequality (\ref{eq:ineq}) to be saturated, it is required that the 1-dimensional unstable manifold along the maximally unstable Lyapunov direction densely covers a finite-measure portion of
the $2d$ dimensional phase space, and moreover, that the exploration of the modes of the Fourier space is not sparse, as is for example in the case of linear automorphisms on the torus ({\em cat maps}).
This may typically be the case - as indicated later - at least in low dimensional maps.

\begin{figure}[h!]
\includegraphics[width=0.8\linewidth]{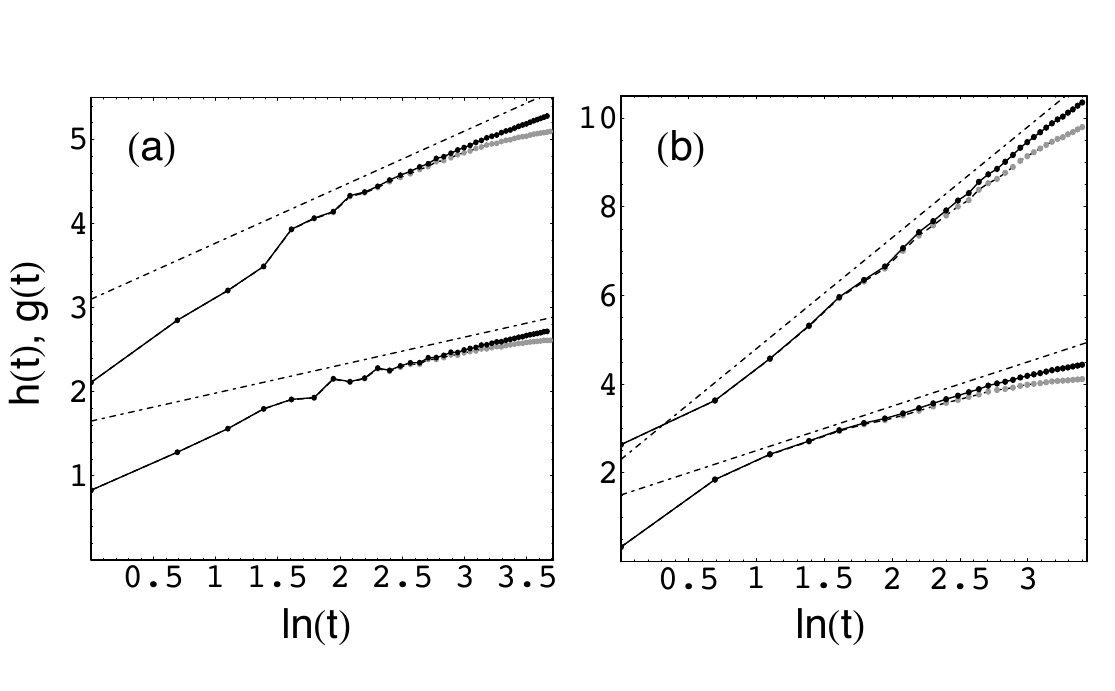}
\caption{Separability entropy $h(t)=h[\rho^t]$ (lower curves) and Fourier entropy $g(t) = g[\rho_t]$ (upper curves) for non-chaotic dynamics, 
integrable IM (a), and non-integrable TM (b). Data for discrectization dimension $N=2^p$ with $p=14$ (black curves), and $p=12$ (grey curves) are shown, where dash-dotted lines suggest asymptotic logarithmic growths $\sim \xi \ln t$, with $\xi=0.333$ (s-entropy for IM), $\xi = 0.667$ (f-entropy for IM), $\xi = 1.0$ (s-entropy for TM), $\xi= 2.5$ (f-entropy for TM).}
\label{fig:eereg} 
\end{figure}

\begin{figure}[h!]
\includegraphics[width=0.9\linewidth]{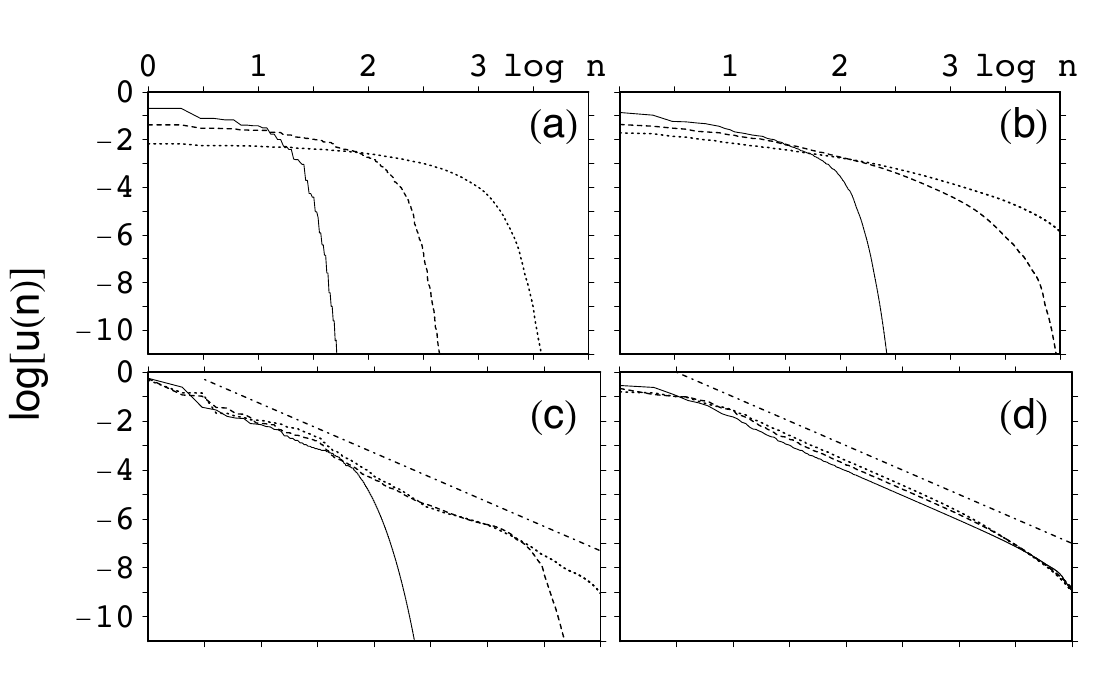}
\caption{Singular value (Schmidt) spectra as a function of time for $t=3$ (full curves), $t=5$ (long dashed), $t=7$ (short dashed), for the dynamics: PC (a), SM (b), IM (c), TM (d) using discretization dimension $N=2^{14}$. $\log u(n)$, with $u(n) = |\mu^t_n|^2$, is plotted against $\log n$, and in non-chaotic cases (c,d), the dash-dotted line indicates $u(n) \propto 1/n^2$ scaling.}
\label{fig:sv} 
\end{figure}

It is interesting to note that for any map with a {\em finite Markov partition} -- and thus admitting exact symbolic dynamics
with a finite grammar -- we have $C_{\rm s}=0$ since in the Markov coordinates the separability (or the number of terms in (\ref{eq:decomp})) is preserved. For linear toral (cat) maps we even have $C_{\rm f} = 0$ since the number of Fourier harmonics is preserved in time
even though their magnitude may be growing. Positive s-compexity, $C_{\rm s} > 0$, thus represents a very strong condition implying practical 
unsolvability of Liouville dynamics due to chaotic motion {\em and} non-existence of a finite Markov partition.

\section{Numerical examples}

Let us now illustrate our concepts by discussing a set of numerical experiments.
Firstly, we consider four different examples of 2D ($d=1$) symplectic toral maps, $(x',y') = \ve{\phi}(x,y)$:
(i) {\em perturbed cat map} (PC) (as in \cite{dana})  $y'=y+x - \alpha\sin x,x'=x+y'$ with $\alpha=0.5$ as an example of non-linear,
non-markovian but uniformly hyperbolic Anosov system,
(ii) {\em non-symmetric standard map} (SM) $y'=y+\alpha\sin x + \beta\cos(2 x), x'=x+y'$ with $\alpha=2,\beta=2$ as an example of strongly chaotic but non-uniformly hyperbolic system with small islands
of regular motion of negligible area,
(iii) {\em integrable (Suris) map} \cite{suris} (IM) $x'=2x+4\arg(1+\alpha e^{-\ii x})-y,y'=x$ with $\alpha=0.5$ as a
 an an example of a non-trivially integrable map with a separatrix,
(iv) {\em triangle map} \cite{tm} (TM) $y'=y + \alpha\,{\rm sgn}(x-\pi) + \beta,x'=x+y'$ with $\alpha = \pi (\sqrt{5}-1)/2,\beta= \pi e^{-1}$
as an example of dynamically mixing system without exponential sensitivity (all
assignments understood ${\rm mod}\; 2\pi$).
In Fig.~\ref{fig:snapsr} we show time evolving phase space densities $\rho^{-t}$ for all four maps
at $t=3,5,7$, all starting from the same simple initial density $\rho^0(x,y) = (2 + \cos x + \cos y)(2\pi\sqrt{5})$.
Note that the three maps PC, SM and TM exhibit dynamical mixing behavior, although for TM the mixing
mechanism is qualitatively different \cite{mix}.
Only the orbits of the first two maps (PC and SM) have positive Kolmogorov complexity, with estimated Lyapunov exponents (being equal to Kolmogorov-Sinai entropies) $\lambda_{\rm max}^{\rm PC} = 0.9496$, $\lambda_{\rm max}^{\rm SM} = 0.8206$,
and only PC exhibits exponential decay of correlations, 
$\int\dd^2\ve{z} \rho^0(\ve{z})\rho^t(\ve{z}) - 1 \sim \exp(-\xi t)$, with
$\xi^{\rm PC} = 1.17$, while for SM and TM correlations decay as power laws.
In Fig.~\ref{fig:snapsf} we display the corresponding Fourier transformed densities 
$\tilde{\rho}^{-t}$ in order to demonstrate the exponential expansion of the distributions of Fourier
harmonics in the chaotic cases (PC,SM), resulting in a positive f-complexity.
Indeed, as we show in Fig.~\ref{fig:eech}, the f-entropy grows linearly with the upper bound
Lyapunov rate (\ref{eq:ineq}), namely $C_{\rm f} = 2 \lambda_{\rm max}$ which we believe should
be a generic behavior for chaotic maps, whereas for s-complexity we find consistently smaller values
$C^{\rm PC}_{\rm s} = 1.00$, $C^{\rm SM}_{\rm s} = 0.952$.
Note that completely different behavior is found for linear chaotic maps, or maps with exact symbolic dynamics like un-perturbed cat map or baker's maps, where we find $C_{\rm s}=0$.
In non-chaotic maps (IM,TM) we find zero s/f-compexity, where the temporal growth of s/f-entropy is
likely to be logarithmic (see Fig.~\ref{fig:eereg} and its caption for details).

The numerical results on s-complexity are supplemented by showing the temporal snapshots of the full Schmidt spectrum $\mu_n^t$ in 
Fig.~\ref{fig:sv}. In the chaotic cases (PC, SM) with positive s-complexity, the full spectrum asymptotically scales as $\mu_n^t \propto f(n/(C_{\rm s} t))$, and the tail of $f(x)$ decays faster than the power law, while in the regular/non-chaotic cases (IM, TM), $\mu_n^t$ converges, 
as $t\to \infty$, to a universal power-law profile $\mu_n^t \to {\rm const}/n$.

\begin{figure}[h!]
\includegraphics[width=0.9\linewidth]{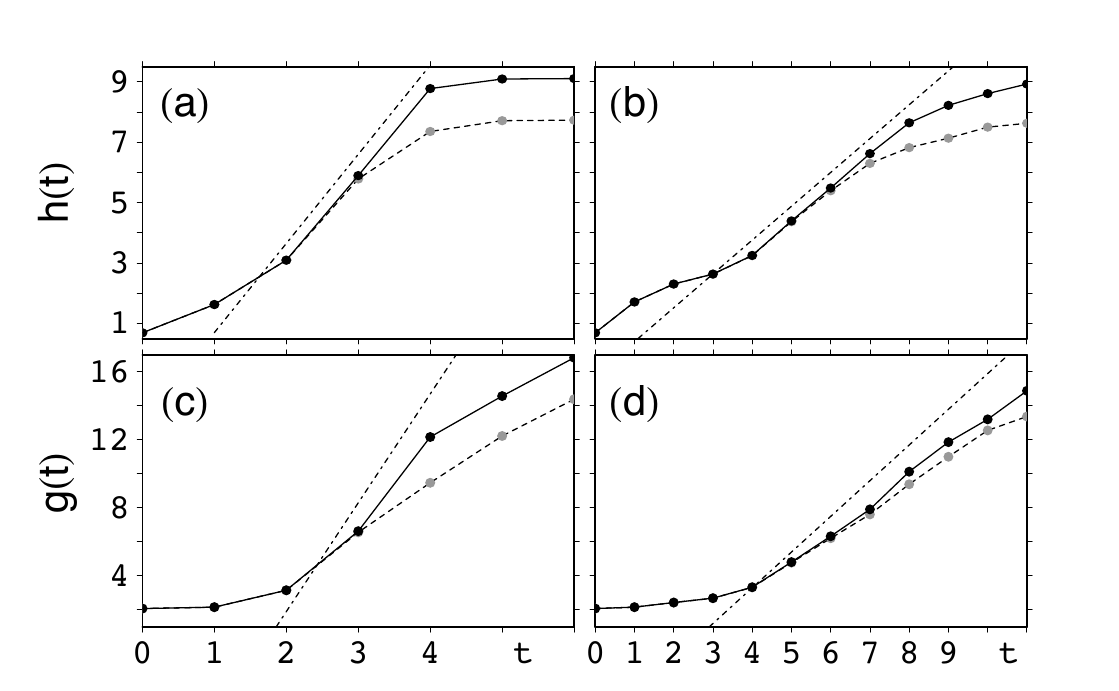}
\caption{Separability entropy $h(t) = h[\rho^t]$ and Fourier entropy $g(t) = g[\rho^t]$ for 4D perturbed cat maps: DH (a) and Lo (b).
We take discretization/truncation to $N=2^p$ nodes in each of 4 phase space directions with $p=7$ (black - full curves) and $p=6$ (gray - dashed curves), initial state described in text. The chain straight lines give the suggested asymptotic rates, $C^{\rm DH}_{\rm s} = 2.95$ (a), $C^{\rm Lo}_{\rm s} = 1.12$ (b), $C^{\rm DH, Lo}_{\rm f} = 4 \lambda^{\rm DH,Lo}_{\rm max}$ (c,d).}
\label{fig:ee4d} 
\end{figure}

Secondly, we consider an example of 4D ($d=2$) toral automorphism, a simple extension of a perturbed cat map to $\T^4$,
 $\ve{\phi}(\ve{z})\equiv (z'_1+\beta_1 \sin z'_3,z'_2+\beta_2 \sin z'_4,z'_3,z'_4)$, 
and $\ve{z}' \equiv {\bf C}\ve{z}$. As for linear part we take exactly the same two cases as in Ref. \cite{veble}, namely the {\em doubly-hyperbolic} (DH), and {\em loxodromic} (Lo) one, with $4\times 4$ matrices
$$
{\bf C}_{\rm DH} = \begin{pmatrix} 2 & -2 & -1 & 0 \cr -2 & 3 & 1 & 0 \cr -1 & 2 & 2 & 1 \cr 2 & -2 & 0 & 1 \end{pmatrix},\quad
{\bf C}_{\rm Lo} = \begin{pmatrix} 0 & 1 & 0 & 0 \cr 0 & 1 & 1 & 0 \cr 1 & -1 & 1 & 1 \cr -1 & -1 & -2 & 0 \end{pmatrix}
$$
and take {\em non-linearities} $\beta_1 = 0.2,\beta_2=0.3$, resulting in, respectively, maximal Lyapunov exponents $\lambda_{\rm max}^{\rm DH} = 1.60$, $\lambda_{\rm max}^{\rm Lo} = 0.525$. Decomposing $\ve{x}=(z_1,z_2),\ve{y}=(z_3,z_4)$, we show in Fig.\ref{fig:ee4d} numerical simulation of s- and f-entropy, starting from the initial state with random lowest Fourier harmonics, i.e. $\tilde{\rho}^0_{\ve{k}}$ being independent random complex Gaussian variables for  $\ve{k}\cdot\ve{k} \le 1$ and $\tilde{\rho}^0_{\ve{k}}=0$ otherwise.
Again, we obtain, consistently with the 2D case, that the f-entropy grows with the rate which is close to $4 \lambda_{\rm max}$, saturating the second bound in (\ref{eq:ineq}), and that the s-complexity is systematically substantially smaller but positive, namely $C^{\rm DH}_{\rm s} = 2.95$, $C^{\rm Lo}_{\rm s} = 1.12$.

It should be noted that our numerical experiments provide only a partial support for the meaningfulness of the definitions and conjectures stated in Section II, although the results seem very suggestive. For example, the supremum over initial density $\rho^0$ has been tested by increasing the Fourier support of $\rho^0$, which typically did not result in appreciable difference in the asymptotic growth rate of $h[\rho^t]$. On the other hand,
we have not yet been able to address systematically the infimum over the phase space partitions $\pi$ in the definition (\ref{eq:scomplex}). However, several trials of varying $\pi$ indicated that the numerical result -- the value of $C_{\rm s}$ -- may indeed be insensitive to composing with (smooth) $\pi$, whereas it seems very plausible that for non-smooth $\pi$ the asymptotic growth rate of $h[\rho^0 \circ \ve{\phi}^{-t} \circ \ve{\pi}]$ cannot lower. Furthermore, it would be a future challenge to come up with analytically solvable examples of s-complex Lyiouville dynamics where the positivity $C_{\rm s} > 0$ could be rigorously proven.

\section{Conclusion}

We have proposed a simple quantitative measure of complexity of classical non-dissipative Liouvillian dynamics. The so-called separability entropy (whose asymptotic growth rate defines what we call s-complexity) is inspired by the entanglement entropy \cite{eisert} 
of quantum states, adapted to classical joint probability distributions of several, or many variables. 
Note that a similar complexity measure in the {\em quantum Liouville space} (or operator space) has been used as an indicator of quantum dynamical complexity and {\em quantum chaos} \cite{PZ07}. It has been argued here that the separability entropy measures the minimal amount of computation resources needed to simulate the classical Liouvillian evolution. Based on simple numerical examples of discrete time dynamical systems on 2D and 4D compact phase space we have demonstrated that s-complexity is non-trivial and typically smaller than the exponential growth rate of the number of Fourier harmonics (f-complexity). For example, for Hamiltonian dynamics with many degrees of freedom one might encounter interesting situations with strong Hamiltonian turbulence (large f-complexity), which may be efficiently simulable by a {\em classical} version of time-dependent density matrix renormalization group  
(a la \cite{vidal}) if s-complexity is small. Our concept is fundamentally different from other popular complexity measures in chaos theory, such as the Kolmogorov-Sinai entropy, which characterize complexity of individual trajectories and often fail to provide any meaningful complexity information
 about the time-dependent Liouvillian density.

Our concepts have interesting quantum extensions.
We note that f-complexity has already been used in order to characterize the complexity of quantum time evolution in terms of a Wigner function \cite{benenti}.
We suggest that s-complexity could have a similar quantum phase space extension, but providing a sharper discriminant between quantum chaotic and quantum regular motions.
\\\\
The author thanks Slovenian research agency, program P1-0044, grant J1-2208, and Alexander von Humboldt fundation for support.
Fruitful discussions by G. Benenti, M. Horvat and M. \v Znidari\v c are gratefully acknowledged.


\begin{thebibliography}{1}
\bibitem{note} Based on numerical experience - some of which is reported later - we conjecture that the asymptotic rate of (\ref{eq:scomplex}) does not even depend on diffeomorphism $\ve{\pi}$. Thus it seems that in practice one can take identity $\ve{\pi}= id$, whereas the supremum over $\rho_0$ is achieved already for very simple nontrivial initial densities (see e.g. Fig. 3).
\end{thebibliography}
\end{document}